\documentclass[aps,prl,reprint,superscriptaddress,floatfix]{revtex4-2}

\usepackage[utf8]{inputenc}
\usepackage{graphicx}
\usepackage{amsmath,amssymb}
\usepackage{siunitx}
\usepackage{hyperref}

\begin{document}

\title{Amorphous Silicates -- Time-Current Superposition and the Dynamics of Plastic Flow in the Glassy State}

\author{Matthieu Bourguignon}
\affiliation{Soft Matter Sciences and Engineering, ESPCI Paris, PSL University, CNRS, Sorbonne University, 75005 Paris, France.}

\author{Gustavo A. Rosales-Sosa}
\author{Yoshinari Kato}
\affiliation{Fundamental Technology Division, Nippon Electric Glass Co., Ltd., 7-1, Seiran 2-Chome, Otsu, 520-8639, Shiga, Japan}

\author{Sergio Sao-Joao}
\author{Morgan Rusinowicz}
\author{Guillaume Kermouche}
\affiliation{Mines Saint-Etienne, Univ Lyon, CNRS, UMR 5307 LGF, Centre SMS, F - 42023 Saint-Etienne France}

\author{Etienne Barthel}
\email{etienne.barthel@espci.fr}
\affiliation{Soft Matter Sciences and Engineering, ESPCI Paris, PSL University, CNRS, Sorbonne University, 75005 Paris, France.}

\date{\today}

\begin{abstract}
Electron irradiation enables quantitative control over the plastic flow dynamics of silicate glasses, even far below the glass transition temperature. Through stress-relaxation experiments spanning ambient to near-glass-transition temperatures, we uncover a time–current equivalence that grants direct access to steady-state plastic flow over five decades in strain rate. This equivalence allows reconstruction of the intrinsic plastic-flow curve and quantitative assessment of the roles of network connectivity and temperature. Notably, the observed temperature dependence reveals a striking discrepancy with existing theoretical frameworks, highlighting the need for a comprehensive model of plastic flow dynamics in the glassy state.
\end{abstract}

\maketitle

Plastic flow in amorphous solids remains poorly understood, particularly deep in the glassy state where deformation localizes into shear bands and frequently culminates in brittle failure. In soft glassy materials, flow dynamics are comparatively well characterized, with established experimental and theoretical frameworks capturing their spatiotemporal response~\cite{Cipelletti00, Goyon08, Divoux10, Nicolas18}. By contrast, stiff amorphous solids, including network glasses such as silicates, bulk metallic glasses and glassy polymers still lack a unified description of plastic flow, especially far below the glass transition temperature $T_g$. In this regime, relaxation times increase dramatically, reaching geological scales, and homogeneous steady-state flow becomes experimentally inaccessible~\cite{Hagan80, Lu03, Gross18}, except close to $T_g$~\cite{Brueckner94, Schuh07} and possibly at very small scales~\cite{Toennies14, Song22}. As a result, direct and quantitative tests of theoretical predictions~\cite{Biroli24, Berthier24} remain extremely limited.

Irradiation offers a distinct pathway to manipulate structural rearrangements in amorphous silicates. Electron beams have been shown to induce nanoscale superplasticity~\cite{Ajayan92, Moebus10, Zheng10, Bruns23}, while hard X-rays accelerate structural relaxation~\cite{Ruta17, Pintori19}. Yet irradiation has not been leveraged to quantitatively investigate plastic-flow dynamics.

In this Letter, we demonstrate that electron irradiation enables quantitative control of plastic flow in silicate glasses, even far below $T_g$. Through stress-relaxation experiments conducted under controlled electron irradiation across temperatures ranging from ambient to near $T_g$, we establish a time–current equivalence that provides direct access to steady-state flow over five decades in strain rate. This equivalence enables reconstruction of the intrinsic plastic-flow curve and reveals a temperature dependence that cannot be reconciled with existing theoretical descriptions. Our findings highlight the need for a more comprehensive framework of plastic flow deep within the glassy state and establish electron irradiation as a powerful experimental handle for probing plasticity in network glasses.

\begin{figure}[]
  \centering
\includegraphics{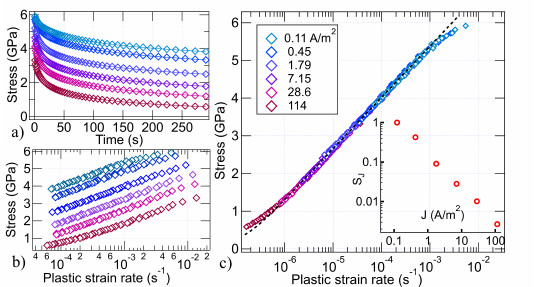}
  \caption{a) Stress relaxation as a function of time measured at increasing electron current density (see panel c for legend), achieved by varying the SEM magnification and/or aperture. Relaxation accelerates systematically with increasing current density. b) Flow curves derived from the relaxation data in panel a). c) Master flow curve obtained by time-rescaling the partial flow curves shown in panel b). Inset: time-scaling factors as a function of current density, revealing an inverse relationship between current density and the characteristic timescale of plastic flow.}\label{Fig:SiO2_rescaling}
\end{figure}
In situ micropillar compression experiments were performed inside a scanning electron microscope (SEM) following the protocol described in our previous work~\cite{Kermouche16}. Three types of silicate glasses were investigated: amorphous silica (SiO$_2$),  calcium aluminoborosilicate (ABS) and calcium soda–silicate (SLS). In brief, micropillars (radius 4~$\mu$m, aspect ratio 1.1) were fabricated by reactive ion etching followed by focused ion beam trimming. The 20~keV electron irradiation during compression was controlled by the current density $J$ (0.1–100~A/m$^2$)~\cite{Rusinowicz25}, ensuring homogeneous irradiation of the pillars~\cite{Bruns23}. Raman spectroscopy indicates that these irradiation conditions leave the final glass structure essentially unchanged (see Supplemental Material for experimental details on the glasses, sample fabrication and Raman analysis).


To quantify the flow dynamics under irradiation, we apply rapid loading to a fixed displacement plateau and record the ensuing stress relaxation over 300~s  (see Supplemental Material for detail). In direct correspondence with previous reports of irradiation-enhanced plasticity~\cite{Zheng10, Bruns23, Rusinowicz25}, we find that higher current densities produce faster and more complete stress relaxation (Fig.~\ref{Fig:SiO2_rescaling} a).

We convert the stress relaxation data into flow curves, that is stress $\sigma$ versus plastic strain rate $\dot{\epsilon}_p$, using $\dot{\epsilon}_p = -\dot{\sigma}/E$ where $E$ is Young's modulus~\cite{Baral19} (Fig.~\ref{Fig:SiO2_rescaling}b). Remarkably, the flow curves obtained at different current densities can be superimposed by a simple rescaling of time
\begin{equation}
dt_I = S_I\, d\tau
\end{equation}
where $S_I$ is a current-dependent scaling factor, $\tau$ the physical time and $t_I$ the scaled time. This procedure mirrors the well-known time–temperature superposition principle used to collapse dynamical data near the glass transition~\cite{Ferry80}. In logarithmic representation, time rescaling corresponds to horizontal shifts of the partial flow curves. Taking the lowest current $I_0$ as the reference, all curves collapse onto a single master curve (Fig.~\ref{Fig:SiO2_rescaling}c), demonstrating a strict equivalence between current density and time.


In contrast to time-temperature superposition, though, we find that in amorphous silica under electron irradiation, the scaling factor follows a power-law dependence $S_I \propto i^{-c}$ with $c \approx 0.87$, close to 1 (Fig.~\ref{Fig:SiO2_rescaling} c, inset). This nearly inverse relationship between current density and characteristic time, which holds over 3 decades of current, leads to a particularly simple conclusion: at fixed applied stress, the plastic strain rate is proportional to the current density.

This scaling can be rationalized within a simple microscopic picture. In an Eyring-type description of plastic flow in amorphous materials~\cite{Spaepen77, Schuh07}, one considers a density $n$ of independent flow sites that undergo stress-assisted transitions between configurations, each event producing a permanent local shear strain $\gamma$ within a characteristic volume $V$. In this picture, the macroscopic plastic strain rate can be expressed as
\begin{equation}\label{eq:plastic strain rate}
\dot{\epsilon} = n \gamma V\ k(T,\sigma)
\end{equation}
where $k$ denotes the flip rate of individual plastic events~\cite{Spaepen77}. Thus, the proportionality between current density and strain rate indicates that the density of plastic flow sites is directly proportional to the current density of the electrons that generate them. At the same time, the master flow curve captures the stress dependence of the intrinsic flip rate $k$ (Fig.~\ref{Fig:SiO2_rescaling}c). Here, we find that it is well described by the functional form
\begin{equation}\label{eq:plastic strain rate k}
\dot{\epsilon}_p = a \sinh(\sigma/b)
\end{equation}
Taken together, these results support a remarkably simple picture of plastic flow deep in the glassy state: electron irradiation controls the density of plastic flow sites, while their individual dynamics remain governed solely by stress and temperature. As an example, the functional form~\ref{eq:plastic strain rate k} is consistent with rate theory~\cite{Spaepen77} where
\begin{equation}\label{eq:plastic strain rate k}
k(T,\sigma) = \nu \exp\left(-\Delta E/kT\right)\sinh\left(\Omega\sigma/kT\right)
\end{equation}
with $\nu$ an attempt frequency of the order of the Debye frequency, $\Delta E$ the activation energy and $\Omega$ the activation volume. 

\begin{figure}[]
  \centering
  \includegraphics{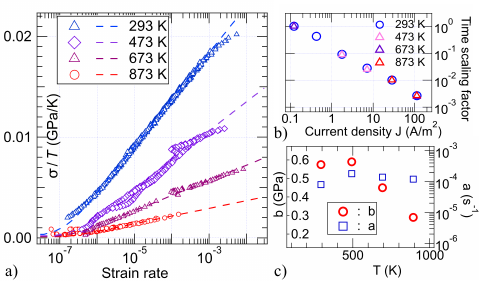}
  \caption{Impact of temperature on the plastic flow of silica under electron irradiation: a) reference flow curves for various temperatures up to 873 K - stress has been normalized to temperature; b) time scaling factors as a function of current - the time scaling factors are independent of temperature; c) flow rule parameters $a$ and $b$as a function of temperature. The process seems largely athermal with little evolution of $a$ and $b$ below 500~K but but a significant decrease at higher temperatures.}\label{Fig:Temperature_dependance}
\end{figure}

To further evaluate the robustness of our approech, we have conducted experiments on amorphous silica at elevated temperatures of 473, 673, and 873~K. Owing to the experimental challenges inherent to high-temperature micromechanical testing, the data exhibit increased scatter. Nevertheless, we find that time–current superposition remains valid across the entire temperature range investigated. The collapsed flow curves resulting from time-current superposition are shown in Fig.~\ref{Fig:Temperature_dependance}a, with the corresponding shift factors presented in Fig.~\ref{Fig:Temperature_dependance}b. Notably, for a given current density, the shift factors are essentially temperature independent, indicating that the mechanism governing the formation of flow sites through electron irradiation is athermal.

In contrast, the flow curves show a strong temperature dependence. To test consistency with rate theory (Eq.~\ref{eq:plastic strain rate k}), data at different temperatures are replotted in Fig.~\ref{Fig:Temperature_dependance}a as $\sigma/T$ versus $\ln(\dot{\epsilon}_p)$, a representation commonly used for glassy polymers~\cite{Bauwens_Crowet_1969}. Within this framework, one expects parallel straight lines with slope set by the activation volume $\Omega$, and intercepts scaling with reciprocal temperature, reflecting the activation energy $\Delta E$. Although some scatter appears at higher temperatures, the data clearly deviate from this expectation: the slopes exhibit a pronounced temperature dependence whereas the intercept remains nearly unchanged. This departure demonstrates that plastic flow in the glassy state cannot be described as a collection of independent, thermally activated flip events, emphasizing the need for a more sophisticated framework to capture the dynamics of flow deep within the glassy state.

%
Building on these insights into the flow dynamics of the archetypal network glass, amorphous silica, we next address the role of network modifications. Network modifiers such as Na$^+$ are known to alter glass dynamics substantially: they lower $T_g$, increase fragility in the supercooled liquid state~\cite{Angell95}, and promote strain localization, including shear bands and shear flaws, below $T_g$~\cite{Arora79}. We examined room-temperature plastic flow in network-modified glasses under identical irradiation conditions. Relaxation experiments were therefore conducted on calcium soda–silicate (SLS, window glass) and calcium aluminoborosilicate (ABS) compositions. Owing to less efficient reactive ion etching, the final pillar geometry was defined by focused ion beam trimming, while all other experimental parameters were kept identical (see Supplemental Material).

\begin{figure}
  \centering
\includegraphics{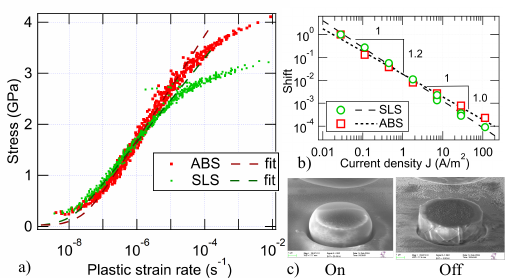}
  \caption{Impact of network depolymerization on the plastic flow of amorphous silicates under electron irradiation: a) master flow curves for SLS and ABS glasses; b) time scaling factors as a function of current density for SLS and ABS. As for silica, the relation is nearly inverse, but the exponents are slightly larger; c) SLS pillar morphologies after compression at constant strain rate under electron irradiation (left) and in the absence of irradiation (right). Shear bands are characteristic of the compression of ABS and SLS but are suppressed under irradiation. 
  }\label{Fig:other_glasses_rescaling}
\end{figure}

In both systems, the time–current equivalence principle remains valid, as demonstrated by the excellent collapse of the flow curves after horizontal shifting at different current densities (Fig.~\ref{Fig:other_glasses_rescaling}a). The quasi-proportionality between current density and strain rate is likewise preserved, evidenced by the nearly inverse scaling of the time shift with current density, although the scaling exponents are slightly larger with c = 1.0 for ABS and c = 1.2 for SLS (Fig.~\ref{Fig:other_glasses_rescaling}b). At low strain rates (high current densities), the flow curves exhibit the same logarithmic strain-rate dependence observed in amorphous silica over the full range of strain rates. However, for SLS and ABS, a crossover emerges at strain rates of order 10$^{-5}$, beyond which a distinct regime with reduced strain-rate sensitivity emerges. At the same time, compression tests performed without irradiation reveal pronounced shear bands in these same glasses (Fig.~\ref{Fig:other_glasses_rescaling}c), whereas amorphous silica shows no such localization under identical conditions, nor do SLS and ABS at high current densities, in line with earlier reports of irradiation-induced suppression of indentation fracture in similar silicate glasses~\cite{Bruns25}. In fact, these results parallel observations near $T_g$ in bulk metallic glasses (BMGs)~\cite{Lu03} and silicate glasses~\cite{Brueckner94}, where strain localization is linked to reduced strain-rate sensitivity, as originally described by Spaepen~\cite{Spaepen77}.


%


In brief, the impact of electron irradiation on plastic flow can be summarized as follows: 1) the characteristic timescale of the flow is inversely proportional to the current density; 2) the effect is transient: it does not depend on irradiation duration, does not induce permanent structural changes, and vanishes once the irradiation ceases; and 3) the process is athermal in nature. Accordingly, stress relaxation under electron irradiation exhibits the same phenomenology as structural relaxation under hard X-ray exposure, as revealed by XPCS measurements~\cite{Ruta17, Pintori19}. 

These observations identify a specific mechanism for irradiation-induced plastic rearrangements, unifying descriptions previously framed as bond switching in irradiation-induced superplasticity~\cite{Zheng10} or as radiolysis in structural relaxation studies~\cite{Ruta17}. Cathodoluminescence measurements show that incident electrons at $\sim$10 keV undergo both elastic and inelastic scattering, generating cascades of low-energy electronic excitations with characteristic energies of order $\sim$10 eV, comparable to the silica band gap~\cite{StevensKalceff13}. This energy scale far exceeds thermal energies over the investigated temperature range, establishing the athermal nature of the process.

These excitations populate antibonding Si–O states at the bottom of the conduction band, transiently weakening bonds and promoting bond rupture, thereby locally reducing network connectivity and enabling plastic rearrangements. The underlying excitation is the self-trapped exciton (STE), extensively characterized by \emph{ab initio} calculations~\cite{IsmailBeigi05a, Kang23}. Radiative recombination of the STE, responsible for the well-known 2.4~eV emission, restores the pristine silica network~\cite{StevensKalceff13}, leaving no permanent defect.

Moreover, STEs exhibit lifetimes of order $\sim$1~$\mu$s at room temperature~\cite{Stathis87}, about seven orders of magnitude longer than the Debye period (Eq.~\ref{eq:plastic strain rate k}). As a result, under electron irradiation STEs act as quasi-static flow defects but they vanish as soon as the beam is turned off.

In brief, the cascade mechanism producing an STE density directly proportional to the current density (Fig.~\ref{Fig:SiO2_rescaling}c inset; Fig.~\ref{Fig:other_glasses_rescaling}b), the high excitation energy ensuring athermal behavior (Fig.~\ref{Fig:Temperature_dependance}b), the reduced connectivity, the intermediate lifetime and the defect-free recombination collectively account for the observed features of irradiation-induced viscoplasticity and support the identification of STEs as the primary electron-induced flow sites.

Resulting from this flow site generation mechanism, the time-current equivalence allows reconstruction of the intrinsic flow-site dynamics in the glassy state over an wide range of strain rates (Fig.~\ref{Fig:SiO2_rescaling}c). While the logarithmic strain-rate dependence of the resulting flow curve is consistent with Eyring-type rate theory~\cite{Spaepen77}, the temperature dependence of plastic flow is not. In particular, the activation parameter $b$ in Eq.~\ref{eq:plastic strain rate} remains nearly constant below 500 K, then decreases, extrapolating to zero at a glass transition temperature $T_g \simeq 1000$ K (Fig.~\ref{Fig:Temperature_dependance}c). This trend closely parallels the temperature dependence of the structural relaxation time under hard X-ray irradiation in B$_2$O$_3$~\cite{Pintori19}. Such behavior is incompatible not only with simple rate theories which predict a $T^{-1}$ dependence, but also with the $(T/T_g)^{2/3}$ scaling of the yield stress proposed by cooperative barrier theory~\cite{Johnson05}. While logarithmic scaling with strain rate arises in phenomenological rate-and-state friction laws~\cite{Dieterich79}, granular dynamics~\cite{Hartley03}, and numerical treatments based on random first-order transition theory~\cite{Wisitsorasak17} or metadynamics~\cite{Cao19}, it is the temperature dependence that constitute the key constraint. This underscores the absence of a comprehensive theoretical framework capable of describing the evolution of plastic flow deep within the glassy state. Our results therefore provide a benchmark against which emerging models of glass plasticity can be evaluated.

Finally, we examine the relationship between network structure and dynamics. Silica glass under irradiation should not be viewed as pristine silica with merely accelerated dynamics, but rather as a distinct glassy state in which self-trapped excitons (STEs) act as transient network modifiers that locally reduce connectivity. This interpretation is supported by the effective glass transition temperature, $T_g \simeq 1000$ K, inferred from the temperature dependence of the activation parameter $b$, which is markedly lower than the $\sim$1500 K characteristic of pure amorphous silica.

This perspective also clarifies the transition from shear localization at high strain rates to homogeneous flow at lower strain rates observed in both SLS and ABS. In these materials, plastic deformation is expected to concentrate around modifier cations, which locally alter network connectivity. Electron irradiation, however, generates an additional population of low–yield-stress flow sites. As their density increases with current, these sites promote more spatially distributed rearrangements, thereby stabilizing homogeneous plastic flow and driving a crossover from localized to homogeneous deformation.

Whether these two populations of flow sites (intrinsic network modifiers and irradiation-induced STEs) act independently or cooperatively remains an open question that calls for more advanced experimental investigation. In this respect, irradiation provides a powerful and tunable means to probe the diffusion of localization as a mechanism for improved resistance to rupture in glassy networks.



In summary, we show that electron irradiation generates transient flow sites in numbers proportional to current, establishing a time–current equivalence that enables reconstruction of intrinsic flow dynamics over an extended range of strain rates deep within the glassy state. While the logarithmic kinetics of the flow curve align broadly with several theoretical models, its temperature dependence reveals significant discrepancies with current theoretical frameworks for the dynamics of glass plasticity.

We identify self-trapped excitons as the primary electron-induced flow sites. Their excitation energy enables athermal activation, their reduced local connectivity facilitates plastic rearrangements, and their defect-free recombination preserves the network’s integrity. Under irradiation, silica should therefore be regarded not merely as pristine glass with accelerated dynamics, but as a distinct glassy state in which transient electronic excitations act as reversible network modifiers. The associated reduction in the effective glass transition temperature further supports this interpretation. Similar principles extend to network-modified glasses: in ABS and SLS, irradiation generates additional flow sites that compete with those introduced by network modifiers, thereby diffusing strain localization and suppressing rupture.

Beyond elucidating the microscopic origins of irradiation-induced plasticity, our results offer a controlled method for tuning flow-site density and probing the complex interplay between structure, dynamics, and localization in glasses. Electron irradiation thus emerges as a powerful experimental tool for investigating the fundamental mechanisms of plastic flow deep in the glassy state, thus setting a benchmark for theoretical models of amorphous plasticity.

\begin{acknowledgments}
This work was supported by Nippon Electric Glass Co., Ltd. and the French Agence Nationale de la Recherche through the GaLAaD (ANR-20-CE08-0002) project. We thank Sylvain Patinet and Fabien Volpi for interesting discussions.
\end{acknowledgments}

\bibliographystyle{apsrev4-1} 
\bibliography{Mechanics_sdrive_2} 

\begin{thebibliography}{37}%
\makeatletter
\providecommand \@ifxundefined [1]{%
 \@ifx{#1\undefined}
}%
\providecommand \@ifnum [1]{%
 \ifnum #1\expandafter \@firstoftwo
 \else \expandafter \@secondoftwo
 \fi
}%
\providecommand \@ifx [1]{%
 \ifx #1\expandafter \@firstoftwo
 \else \expandafter \@secondoftwo
 \fi
}%
\providecommand \natexlab [1]{#1}%
\providecommand \enquote  [1]{``#1''}%
\providecommand \bibnamefont  [1]{#1}%
\providecommand \bibfnamefont [1]{#1}%
\providecommand \citenamefont [1]{#1}%
\providecommand \href@noop [0]{\@secondoftwo}%
\providecommand \href [0]{\begingroup \@sanitize@url \@href}%
\providecommand \@href[1]{\@@startlink{#1}\@@href}%
\providecommand \@@href[1]{\endgroup#1\@@endlink}%
\providecommand \@sanitize@url [0]{\catcode `\\12\catcode `\$12\catcode `\&12\catcode `\#12\catcode `\^12\catcode `\_12\catcode `\%12\relax}%
\providecommand \@@startlink[1]{}%
\providecommand \@@endlink[0]{}%
\providecommand \url  [0]{\begingroup\@sanitize@url \@url }%
\providecommand \@url [1]{\endgroup\@href {#1}{\urlprefix }}%
\providecommand \urlprefix  [0]{URL }%
\providecommand \Eprint [0]{\href }%
\providecommand \doibase [0]{http://dx.doi.org/}%
\providecommand \selectlanguage [0]{\@gobble}%
\providecommand \bibinfo  [0]{\@secondoftwo}%
\providecommand \bibfield  [0]{\@secondoftwo}%
\providecommand \translation [1]{[#1]}%
\providecommand \BibitemOpen [0]{}%
\providecommand \bibitemStop [0]{}%
\providecommand \bibitemNoStop [0]{.\EOS\space}%
\providecommand \EOS [0]{\spacefactor3000\relax}%
\providecommand \BibitemShut  [1]{\csname bibitem#1\endcsname}%
\let\auto@bib@innerbib\@empty
\bibitem [{\citenamefont {Cipelletti}\ \emph {et~al.}(2000)\citenamefont {Cipelletti}, \citenamefont {Manley}, \citenamefont {Ball},\ and\ \citenamefont {Weitz}}]{Cipelletti00}%
  \BibitemOpen
  \bibfield  {author} {\bibinfo {author} {\bibfnamefont {L.}~\bibnamefont {Cipelletti}}, \bibinfo {author} {\bibfnamefont {S.}~\bibnamefont {Manley}}, \bibinfo {author} {\bibfnamefont {R.~C.}\ \bibnamefont {Ball}}, \ and\ \bibinfo {author} {\bibfnamefont {D.~A.}\ \bibnamefont {Weitz}},\ }\href {\doibase 10.1103/physrevlett.84.2275} {\bibfield  {journal} {\bibinfo  {journal} {Physical Review Letters}\ }\textbf {\bibinfo {volume} {84}},\ \bibinfo {pages} {2275} (\bibinfo {year} {2000})}\BibitemShut {NoStop}%
\bibitem [{\citenamefont {Goyon}\ \emph {et~al.}(2008)\citenamefont {Goyon}, \citenamefont {Colin}, \citenamefont {Ovarlez}, \citenamefont {Ajdari},\ and\ \citenamefont {Bocquet}}]{Goyon08}%
  \BibitemOpen
  \bibfield  {author} {\bibinfo {author} {\bibfnamefont {J.}~\bibnamefont {Goyon}}, \bibinfo {author} {\bibfnamefont {A.}~\bibnamefont {Colin}}, \bibinfo {author} {\bibfnamefont {G.}~\bibnamefont {Ovarlez}}, \bibinfo {author} {\bibfnamefont {A.}~\bibnamefont {Ajdari}}, \ and\ \bibinfo {author} {\bibfnamefont {L.}~\bibnamefont {Bocquet}},\ }\href {\doibase 10.1038/nature07026} {\bibfield  {journal} {\bibinfo  {journal} {Nature}\ }\textbf {\bibinfo {volume} {454}},\ \bibinfo {pages} {84} (\bibinfo {year} {2008})}\BibitemShut {NoStop}%
\bibitem [{\citenamefont {Divoux}\ \emph {et~al.}(2010)\citenamefont {Divoux}, \citenamefont {Tamarii}, \citenamefont {Barentin},\ and\ \citenamefont {Manneville}}]{Divoux10}%
  \BibitemOpen
  \bibfield  {author} {\bibinfo {author} {\bibfnamefont {T.}~\bibnamefont {Divoux}}, \bibinfo {author} {\bibfnamefont {D.}~\bibnamefont {Tamarii}}, \bibinfo {author} {\bibfnamefont {C.}~\bibnamefont {Barentin}}, \ and\ \bibinfo {author} {\bibfnamefont {S.}~\bibnamefont {Manneville}},\ }\href {\doibase 10.1103/physrevlett.104.208301} {\bibfield  {journal} {\bibinfo  {journal} {Physical Review Letters}\ }\textbf {\bibinfo {volume} {104}},\ \bibinfo {pages} {208301} (\bibinfo {year} {2010})}\BibitemShut {NoStop}%
\bibitem [{\citenamefont {Nicolas}\ \emph {et~al.}(2018)\citenamefont {Nicolas}, \citenamefont {Ferrero}, \citenamefont {Martens},\ and\ \citenamefont {Barrat}}]{Nicolas18}%
  \BibitemOpen
  \bibfield  {author} {\bibinfo {author} {\bibfnamefont {A.}~\bibnamefont {Nicolas}}, \bibinfo {author} {\bibfnamefont {E.}~\bibnamefont {Ferrero}}, \bibinfo {author} {\bibfnamefont {K.}~\bibnamefont {Martens}}, \ and\ \bibinfo {author} {\bibfnamefont {J.-L.}\ \bibnamefont {Barrat}},\ }\href {\doibase 10.1103/RevModPhys.90.045006} {\bibfield  {journal} {\bibinfo  {journal} {Reviews of Modern Physics}\ }\textbf {\bibinfo {volume} {90}},\ \bibinfo {pages} {045006} (\bibinfo {year} {2018})},\ \Eprint {http://arxiv.org/abs/arXiv:1708.09194v4[cond-mat.dis-nn]} {arXiv:1708.09194v4[cond-mat.dis-nn]} \BibitemShut {NoStop}%
\bibitem [{\citenamefont {Hagan}(1980)}]{Hagan80}%
  \BibitemOpen
  \bibfield  {author} {\bibinfo {author} {\bibfnamefont {J.~T.}\ \bibnamefont {Hagan}},\ }\href {\doibase 10.1007/BF00752121} {\bibfield  {journal} {\bibinfo  {journal} {J. Mater. Sci.}\ }\textbf {\bibinfo {volume} {15}},\ \bibinfo {pages} {1417 } (\bibinfo {year} {1980})}\BibitemShut {NoStop}%
\bibitem [{\citenamefont {Lu}\ \emph {et~al.}(2003)\citenamefont {Lu}, \citenamefont {Ravichandran},\ and\ \citenamefont {Johnson}}]{Lu03}%
  \BibitemOpen
  \bibfield  {author} {\bibinfo {author} {\bibfnamefont {J.}~\bibnamefont {Lu}}, \bibinfo {author} {\bibfnamefont {G.}~\bibnamefont {Ravichandran}}, \ and\ \bibinfo {author} {\bibfnamefont {W.}~\bibnamefont {Johnson}},\ }\href {\doibase 10.1016/S1359-6454(03)00164-2} {\bibfield  {journal} {\bibinfo  {journal} {Acta Materialia}\ }\textbf {\bibinfo {volume} {51}},\ \bibinfo {pages} {3429} (\bibinfo {year} {2003})}\BibitemShut {NoStop}%
\bibitem [{\citenamefont {Gross}\ \emph {et~al.}(2018)\citenamefont {Gross}, \citenamefont {Wu}, \citenamefont {Baker}, \citenamefont {Price},\ and\ \citenamefont {Yongsunthon}}]{Gross18}%
  \BibitemOpen
  \bibfield  {author} {\bibinfo {author} {\bibfnamefont {T.}~\bibnamefont {Gross}}, \bibinfo {author} {\bibfnamefont {J.}~\bibnamefont {Wu}}, \bibinfo {author} {\bibfnamefont {D.}~\bibnamefont {Baker}}, \bibinfo {author} {\bibfnamefont {J.}~\bibnamefont {Price}}, \ and\ \bibinfo {author} {\bibfnamefont {R.}~\bibnamefont {Yongsunthon}},\ }\href@noop {} {\bibfield  {journal} {\bibinfo  {journal} {Journal of Non-Crystalline Solids}\ }\textbf {\bibinfo {volume} {494}},\ \bibinfo {pages} {13} (\bibinfo {year} {2018})}\BibitemShut {NoStop}%
\bibitem [{\citenamefont {Brückner}\ and\ \citenamefont {Yue}(1994)}]{Brueckner94}%
  \BibitemOpen
  \bibfield  {author} {\bibinfo {author} {\bibfnamefont {R.}~\bibnamefont {Brückner}}\ and\ \bibinfo {author} {\bibfnamefont {Y.}~\bibnamefont {Yue}},\ }\href {\doibase 10.1016/0022-3093(94)90003-5} {\bibfield  {journal} {\bibinfo  {journal} {Journal of Non-Crystalline Solids}\ }\textbf {\bibinfo {volume} {175}},\ \bibinfo {pages} {118} (\bibinfo {year} {1994})}\BibitemShut {NoStop}%
\bibitem [{\citenamefont {Schuh}\ \emph {et~al.}(2007)\citenamefont {Schuh}, \citenamefont {Hufnagel},\ and\ \citenamefont {Ramamurty}}]{Schuh07}%
  \BibitemOpen
  \bibfield  {author} {\bibinfo {author} {\bibfnamefont {C.}~\bibnamefont {Schuh}}, \bibinfo {author} {\bibfnamefont {T.}~\bibnamefont {Hufnagel}}, \ and\ \bibinfo {author} {\bibfnamefont {U.}~\bibnamefont {Ramamurty}},\ }\href {\doibase 10.1016/j.actamat.2007.01.052} {\bibfield  {journal} {\bibinfo  {journal} {Acta Materialia}\ }\textbf {\bibinfo {volume} {55}},\ \bibinfo {pages} {4067} (\bibinfo {year} {2007})}\BibitemShut {NoStop}%
\bibitem [{\citenamefont {Tönnies}\ \emph {et~al.}(2014)\citenamefont {Tönnies}, \citenamefont {Maaß},\ and\ \citenamefont {Volkert}}]{Toennies14}%
  \BibitemOpen
  \bibfield  {author} {\bibinfo {author} {\bibfnamefont {D.}~\bibnamefont {Tönnies}}, \bibinfo {author} {\bibfnamefont {R.}~\bibnamefont {Maaß}}, \ and\ \bibinfo {author} {\bibfnamefont {C.~A.}\ \bibnamefont {Volkert}},\ }\href {\doibase 10.1002/adma.201401123} {\bibfield  {journal} {\bibinfo  {journal} {Advanced Materials}\ }\textbf {\bibinfo {volume} {26}},\ \bibinfo {pages} {5715} (\bibinfo {year} {2014})}\BibitemShut {NoStop}%
\bibitem [{\citenamefont {Song}\ \emph {et~al.}(2022)\citenamefont {Song}, \citenamefont {Zhu},\ and\ \citenamefont {Chen}}]{Song22}%
  \BibitemOpen
  \bibfield  {author} {\bibinfo {author} {\bibfnamefont {S.}~\bibnamefont {Song}}, \bibinfo {author} {\bibfnamefont {F.}~\bibnamefont {Zhu}}, \ and\ \bibinfo {author} {\bibfnamefont {M.}~\bibnamefont {Chen}},\ }\href {\doibase 10.1038/s41563-021-01185-y} {\bibfield  {journal} {\bibinfo  {journal} {Nature Materials}\ }\textbf {\bibinfo {volume} {21}},\ \bibinfo {pages} {404} (\bibinfo {year} {2022})}\BibitemShut {NoStop}%
\bibitem [{\citenamefont {Biroli}\ and\ \citenamefont {Bouchaud}(2024)}]{Biroli24}%
  \BibitemOpen
  \bibfield  {author} {\bibinfo {author} {\bibfnamefont {G.}~\bibnamefont {Biroli}}\ and\ \bibinfo {author} {\bibfnamefont {J.-P.}\ \bibnamefont {Bouchaud}},\ }\href {\doibase 10.5802/crphys.136} {\bibfield  {journal} {\bibinfo  {journal} {Comptes Rendus. Physique}\ }\textbf {\bibinfo {volume} {24}},\ \bibinfo {pages} {9} (\bibinfo {year} {2024})}\BibitemShut {NoStop}%
\bibitem [{\citenamefont {Berthier}\ \emph {et~al.}(2025)\citenamefont {Berthier}, \citenamefont {Biroli}, \citenamefont {Manning},\ and\ \citenamefont {Zamponi}}]{Berthier24}%
  \BibitemOpen
  \bibfield  {author} {\bibinfo {author} {\bibfnamefont {L.}~\bibnamefont {Berthier}}, \bibinfo {author} {\bibfnamefont {G.}~\bibnamefont {Biroli}}, \bibinfo {author} {\bibfnamefont {L.}~\bibnamefont {Manning}}, \ and\ \bibinfo {author} {\bibfnamefont {F.}~\bibnamefont {Zamponi}},\ }\href {\doibase 10.1038/s42254-025-00833-5} {\bibfield  {journal} {\bibinfo  {journal} {Nature Reviews Physics}\ }\textbf {\bibinfo {volume} {7}},\ \bibinfo {pages} {313} (\bibinfo {year} {2025})},\ \Eprint {http://arxiv.org/abs/arXiv:2401.09385v1[cond-mat.stat-mech]} {arXiv:2401.09385v1[cond-mat.stat-mech]} \BibitemShut {NoStop}%
\bibitem [{\citenamefont {Ajayan}\ and\ \citenamefont {Iijima}(1992)}]{Ajayan92}%
  \BibitemOpen
  \bibfield  {author} {\bibinfo {author} {\bibfnamefont {P.~M.}\ \bibnamefont {Ajayan}}\ and\ \bibinfo {author} {\bibfnamefont {S.}~\bibnamefont {Iijima}},\ }\href {\doibase 10.1080/09500839208215146} {\bibfield  {journal} {\bibinfo  {journal} {Philosophical Magazine Letters}\ }\textbf {\bibinfo {volume} {65}},\ \bibinfo {pages} {43} (\bibinfo {year} {1992})}\BibitemShut {NoStop}%
\bibitem [{\citenamefont {Möbus}\ \emph {et~al.}(2010)\citenamefont {Möbus}, \citenamefont {Ojovan}, \citenamefont {Cook}, \citenamefont {Tsai},\ and\ \citenamefont {Yang}}]{Moebus10}%
  \BibitemOpen
  \bibfield  {author} {\bibinfo {author} {\bibfnamefont {G.}~\bibnamefont {Möbus}}, \bibinfo {author} {\bibfnamefont {M.}~\bibnamefont {Ojovan}}, \bibinfo {author} {\bibfnamefont {S.}~\bibnamefont {Cook}}, \bibinfo {author} {\bibfnamefont {J.}~\bibnamefont {Tsai}}, \ and\ \bibinfo {author} {\bibfnamefont {G.}~\bibnamefont {Yang}},\ }\href {\doibase 10.1016/j.jnucmat.2009.11.020} {\bibfield  {journal} {\bibinfo  {journal} {Journal of Nuclear Materials}\ }\textbf {\bibinfo {volume} {396}},\ \bibinfo {pages} {264} (\bibinfo {year} {2010})}\BibitemShut {NoStop}%
\bibitem [{\citenamefont {Zheng}\ \emph {et~al.}(2010)\citenamefont {Zheng}, \citenamefont {Wang}, \citenamefont {Cheng}, \citenamefont {Yue}, \citenamefont {Han}, \citenamefont {Zhang}, \citenamefont {Shan}, \citenamefont {Mao}, \citenamefont {Ye}, \citenamefont {Yin},\ and\ \citenamefont {Ma}}]{Zheng10}%
  \BibitemOpen
  \bibfield  {author} {\bibinfo {author} {\bibfnamefont {K.}~\bibnamefont {Zheng}}, \bibinfo {author} {\bibfnamefont {C.}~\bibnamefont {Wang}}, \bibinfo {author} {\bibfnamefont {Y.-Q.}\ \bibnamefont {Cheng}}, \bibinfo {author} {\bibfnamefont {Y.}~\bibnamefont {Yue}}, \bibinfo {author} {\bibfnamefont {X.}~\bibnamefont {Han}}, \bibinfo {author} {\bibfnamefont {Z.}~\bibnamefont {Zhang}}, \bibinfo {author} {\bibfnamefont {Z.}~\bibnamefont {Shan}}, \bibinfo {author} {\bibfnamefont {S.~X.}\ \bibnamefont {Mao}}, \bibinfo {author} {\bibfnamefont {M.}~\bibnamefont {Ye}}, \bibinfo {author} {\bibfnamefont {Y.}~\bibnamefont {Yin}}, \ and\ \bibinfo {author} {\bibfnamefont {E.}~\bibnamefont {Ma}},\ }\href {\doibase 10.1038/ncomms1021} {\bibfield  {journal} {\bibinfo  {journal} {Nature Communications}\ }\textbf {\bibinfo {volume} {1}} (\bibinfo {year} {2010}),\ 10.1038/ncomms1021}\BibitemShut {NoStop}%
\bibitem [{\citenamefont {Bruns}\ \emph {et~al.}(2023)\citenamefont {Bruns}, \citenamefont {Minnert}, \citenamefont {Pethö}, \citenamefont {Michler},\ and\ \citenamefont {Durst}}]{Bruns23}%
  \BibitemOpen
  \bibfield  {author} {\bibinfo {author} {\bibfnamefont {S.}~\bibnamefont {Bruns}}, \bibinfo {author} {\bibfnamefont {C.}~\bibnamefont {Minnert}}, \bibinfo {author} {\bibfnamefont {L.}~\bibnamefont {Pethö}}, \bibinfo {author} {\bibfnamefont {J.}~\bibnamefont {Michler}}, \ and\ \bibinfo {author} {\bibfnamefont {K.}~\bibnamefont {Durst}},\ }\href {\doibase 10.1002/advs.202205237} {\bibfield  {journal} {\bibinfo  {journal} {Advanced Science}\ }\textbf {\bibinfo {volume} {10}} (\bibinfo {year} {2023}),\ 10.1002/advs.202205237}\BibitemShut {NoStop}%
\bibitem [{\citenamefont {Ruta}\ \emph {et~al.}(2017)\citenamefont {Ruta}, \citenamefont {Zontone}, \citenamefont {Chushkin}, \citenamefont {Baldi}, \citenamefont {Pintori}, \citenamefont {Monaco}, \citenamefont {Rufflé},\ and\ \citenamefont {Kob}}]{Ruta17}%
  \BibitemOpen
  \bibfield  {author} {\bibinfo {author} {\bibfnamefont {B.}~\bibnamefont {Ruta}}, \bibinfo {author} {\bibfnamefont {F.}~\bibnamefont {Zontone}}, \bibinfo {author} {\bibfnamefont {Y.}~\bibnamefont {Chushkin}}, \bibinfo {author} {\bibfnamefont {G.}~\bibnamefont {Baldi}}, \bibinfo {author} {\bibfnamefont {G.}~\bibnamefont {Pintori}}, \bibinfo {author} {\bibfnamefont {G.}~\bibnamefont {Monaco}}, \bibinfo {author} {\bibfnamefont {B.}~\bibnamefont {Rufflé}}, \ and\ \bibinfo {author} {\bibfnamefont {W.}~\bibnamefont {Kob}},\ }\href {\doibase 10.1038/s41598-017-04271-x} {\bibfield  {journal} {\bibinfo  {journal} {Scientific Reports}\ }\textbf {\bibinfo {volume} {7}} (\bibinfo {year} {2017}),\ 10.1038/s41598-017-04271-x}\BibitemShut {NoStop}%
\bibitem [{\citenamefont {Pintori}\ \emph {et~al.}(2019)\citenamefont {Pintori}, \citenamefont {Baldi}, \citenamefont {Ruta},\ and\ \citenamefont {Monaco}}]{Pintori19}%
  \BibitemOpen
  \bibfield  {author} {\bibinfo {author} {\bibfnamefont {G.}~\bibnamefont {Pintori}}, \bibinfo {author} {\bibfnamefont {G.}~\bibnamefont {Baldi}}, \bibinfo {author} {\bibfnamefont {B.}~\bibnamefont {Ruta}}, \ and\ \bibinfo {author} {\bibfnamefont {G.}~\bibnamefont {Monaco}},\ }\href {\doibase 10.1103/physrevb.99.224206} {\bibfield  {journal} {\bibinfo  {journal} {Physical Review B}\ }\textbf {\bibinfo {volume} {99}},\ \bibinfo {pages} {224206} (\bibinfo {year} {2019})}\BibitemShut {NoStop}%
\bibitem [{\citenamefont {Kermouche}\ \emph {et~al.}(2016)\citenamefont {Kermouche}, \citenamefont {Guillonneau}, \citenamefont {Michler}, \citenamefont {Teisseire},\ and\ \citenamefont {Barthel}}]{Kermouche16}%
  \BibitemOpen
  \bibfield  {author} {\bibinfo {author} {\bibfnamefont {G.}~\bibnamefont {Kermouche}}, \bibinfo {author} {\bibfnamefont {G.}~\bibnamefont {Guillonneau}}, \bibinfo {author} {\bibfnamefont {J.}~\bibnamefont {Michler}}, \bibinfo {author} {\bibfnamefont {J.}~\bibnamefont {Teisseire}}, \ and\ \bibinfo {author} {\bibfnamefont {E.}~\bibnamefont {Barthel}},\ }\href {\doibase 10.1016/j.actamat.2016.05.027} {\bibfield  {journal} {\bibinfo  {journal} {Acta Materialia}\ }\textbf {\bibinfo {volume} {114}},\ \bibinfo {pages} {146} (\bibinfo {year} {2016})}\BibitemShut {NoStop}%
\bibitem [{\citenamefont {Rusinowicz}\ \emph {et~al.}(2025)\citenamefont {Rusinowicz}, \citenamefont {Sao-Joao}, \citenamefont {Bourguignon}, \citenamefont {Rosales-Sosa}, \citenamefont {Kato}, \citenamefont {Volpi}, \citenamefont {Barthel},\ and\ \citenamefont {Kermouche}}]{Rusinowicz25}%
  \BibitemOpen
  \bibfield  {author} {\bibinfo {author} {\bibfnamefont {M.}~\bibnamefont {Rusinowicz}}, \bibinfo {author} {\bibfnamefont {S.}~\bibnamefont {Sao-Joao}}, \bibinfo {author} {\bibfnamefont {M.}~\bibnamefont {Bourguignon}}, \bibinfo {author} {\bibfnamefont {G.}~\bibnamefont {Rosales-Sosa}}, \bibinfo {author} {\bibfnamefont {Y.}~\bibnamefont {Kato}}, \bibinfo {author} {\bibfnamefont {F.}~\bibnamefont {Volpi}}, \bibinfo {author} {\bibfnamefont {E.}~\bibnamefont {Barthel}}, \ and\ \bibinfo {author} {\bibfnamefont {G.}~\bibnamefont {Kermouche}},\ }\href {\doibase 10.1016/j.scriptamat.2025.116628} {\bibfield  {journal} {\bibinfo  {journal} {Scripta Materialia}\ }\textbf {\bibinfo {volume} {261}},\ \bibinfo {pages} {116628} (\bibinfo {year} {2025})}\BibitemShut {NoStop}%
\bibitem [{\citenamefont {Baral}\ \emph {et~al.}(2019)\citenamefont {Baral}, \citenamefont {Guillonneau}, \citenamefont {Kermouche}, \citenamefont {Bergheau},\ and\ \citenamefont {Loubet}}]{Baral19}%
  \BibitemOpen
  \bibfield  {author} {\bibinfo {author} {\bibfnamefont {P.}~\bibnamefont {Baral}}, \bibinfo {author} {\bibfnamefont {G.}~\bibnamefont {Guillonneau}}, \bibinfo {author} {\bibfnamefont {G.}~\bibnamefont {Kermouche}}, \bibinfo {author} {\bibfnamefont {J.-M.}\ \bibnamefont {Bergheau}}, \ and\ \bibinfo {author} {\bibfnamefont {J.-L.}\ \bibnamefont {Loubet}},\ }\href {\doibase 10.1016/j.mechmat.2019.103095} {\bibfield  {journal} {\bibinfo  {journal} {Mechanics of Materials}\ }\textbf {\bibinfo {volume} {137}},\ \bibinfo {pages} {103095} (\bibinfo {year} {2019})}\BibitemShut {NoStop}%
\bibitem [{\citenamefont {Ferry}(1980)}]{Ferry80}%
  \BibitemOpen
  \bibfield  {author} {\bibinfo {author} {\bibfnamefont {J.~D.}\ \bibnamefont {Ferry}},\ }\href@noop {} {\emph {\bibinfo {title} {Viscoelastic properties of polymers}}}\ (\bibinfo  {publisher} {John Wiley \& Sons},\ \bibinfo {year} {1980})\BibitemShut {NoStop}%
\bibitem [{\citenamefont {Spaepen}(1977)}]{Spaepen77}%
  \BibitemOpen
  \bibfield  {author} {\bibinfo {author} {\bibfnamefont {F.}~\bibnamefont {Spaepen}},\ }\href {\doibase 10.1016/0001-6160(77)90232-2} {\bibfield  {journal} {\bibinfo  {journal} {Acta Metallurgica}\ }\textbf {\bibinfo {volume} {25}},\ \bibinfo {pages} {407} (\bibinfo {year} {1977})}\BibitemShut {NoStop}%
\bibitem [{\citenamefont {Bauwens-Crowet}\ \emph {et~al.}(1969)\citenamefont {Bauwens-Crowet}, \citenamefont {Bauwens},\ and\ \citenamefont {Homès}}]{Bauwens_Crowet_1969}%
  \BibitemOpen
  \bibfield  {author} {\bibinfo {author} {\bibfnamefont {C.}~\bibnamefont {Bauwens-Crowet}}, \bibinfo {author} {\bibfnamefont {J.~C.}\ \bibnamefont {Bauwens}}, \ and\ \bibinfo {author} {\bibfnamefont {G.}~\bibnamefont {Homès}},\ }\href {\doibase 10.1002/pol.1969.160070411} {\bibfield  {journal} {\bibinfo  {journal} {Journal of Polymer Science Part A-2: Polymer Physics}\ }\textbf {\bibinfo {volume} {7}},\ \bibinfo {pages} {735} (\bibinfo {year} {1969})}\BibitemShut {NoStop}%
\bibitem [{\citenamefont {Angell}(1995)}]{Angell95}%
  \BibitemOpen
  \bibfield  {author} {\bibinfo {author} {\bibfnamefont {C.~A.}\ \bibnamefont {Angell}},\ }\href {\doibase 10.1126/science.267.5206.1924} {\bibfield  {journal} {\bibinfo  {journal} {Science}\ }\textbf {\bibinfo {volume} {267}},\ \bibinfo {pages} {1924} (\bibinfo {year} {1995})}\BibitemShut {NoStop}%
\bibitem [{\citenamefont {Arora}\ \emph {et~al.}(1979)\citenamefont {Arora}, \citenamefont {Marshall}, \citenamefont {Lawn},\ and\ \citenamefont {Swain}}]{Arora79}%
  \BibitemOpen
  \bibfield  {author} {\bibinfo {author} {\bibfnamefont {A.}~\bibnamefont {Arora}}, \bibinfo {author} {\bibfnamefont {D.}~\bibnamefont {Marshall}}, \bibinfo {author} {\bibfnamefont {B.}~\bibnamefont {Lawn}}, \ and\ \bibinfo {author} {\bibfnamefont {M.}~\bibnamefont {Swain}},\ }\href@noop {} {\bibfield  {journal} {\bibinfo  {journal} {J. Non-Cryst. Solids}\ }\textbf {\bibinfo {volume} {31}},\ \bibinfo {pages} {415} (\bibinfo {year} {1979})}\BibitemShut {NoStop}%
\bibitem [{\citenamefont {Bruns}\ and\ \citenamefont {Durst}(2025)}]{Bruns25}%
  \BibitemOpen
  \bibfield  {author} {\bibinfo {author} {\bibfnamefont {S.}~\bibnamefont {Bruns}}\ and\ \bibinfo {author} {\bibfnamefont {K.}~\bibnamefont {Durst}},\ }\href@noop {} {\bibfield  {journal} {\bibinfo  {journal} {Materials \& Design}\ }\textbf {\bibinfo {volume} {251}},\ \bibinfo {pages} {113726} (\bibinfo {year} {2025})}\BibitemShut {NoStop}%
\bibitem [{\citenamefont {Stevens-Kalceff}(2013)}]{StevensKalceff13}%
  \BibitemOpen
  \bibfield  {author} {\bibinfo {author} {\bibfnamefont {M.~A.}\ \bibnamefont {Stevens-Kalceff}},\ }\href {\doibase 10.1007/s00710-013-0275-5} {\bibfield  {journal} {\bibinfo  {journal} {Mineralogy and Petrology}\ }\textbf {\bibinfo {volume} {107}},\ \bibinfo {pages} {455} (\bibinfo {year} {2013})}\BibitemShut {NoStop}%
\bibitem [{\citenamefont {Ismail-Beigi}\ and\ \citenamefont {Louie}(2005)}]{IsmailBeigi05a}%
  \BibitemOpen
  \bibfield  {author} {\bibinfo {author} {\bibfnamefont {S.}~\bibnamefont {Ismail-Beigi}}\ and\ \bibinfo {author} {\bibfnamefont {S.~G.}\ \bibnamefont {Louie}},\ }\href {\doibase 10.1103/physrevlett.95.156401} {\bibfield  {journal} {\bibinfo  {journal} {Physical Review Letters}\ }\textbf {\bibinfo {volume} {95}},\ \bibinfo {pages} {156401} (\bibinfo {year} {2005})}\BibitemShut {NoStop}%
\bibitem [{\citenamefont {Kang}\ \emph {et~al.}(2023)\citenamefont {Kang}, \citenamefont {Jeong}, \citenamefont {Paeng}, \citenamefont {Kim}, \citenamefont {Lee}, \citenamefont {Park}, \citenamefont {Han}, \citenamefont {Nam~Han},\ and\ \citenamefont {Choi}}]{Kang23}%
  \BibitemOpen
  \bibfield  {author} {\bibinfo {author} {\bibfnamefont {S.-G.}\ \bibnamefont {Kang}}, \bibinfo {author} {\bibfnamefont {W.}~\bibnamefont {Jeong}}, \bibinfo {author} {\bibfnamefont {J.}~\bibnamefont {Paeng}}, \bibinfo {author} {\bibfnamefont {H.}~\bibnamefont {Kim}}, \bibinfo {author} {\bibfnamefont {E.}~\bibnamefont {Lee}}, \bibinfo {author} {\bibfnamefont {G.-S.}\ \bibnamefont {Park}}, \bibinfo {author} {\bibfnamefont {S.}~\bibnamefont {Han}}, \bibinfo {author} {\bibfnamefont {H.}~\bibnamefont {Nam~Han}}, \ and\ \bibinfo {author} {\bibfnamefont {I.-S.}\ \bibnamefont {Choi}},\ }\href {\doibase 10.1016/j.mattod.2023.04.009} {\bibfield  {journal} {\bibinfo  {journal} {Materials Today}\ }\textbf {\bibinfo {volume} {66}},\ \bibinfo {pages} {62} (\bibinfo {year} {2023})}\BibitemShut {NoStop}%
\bibitem [{\citenamefont {Stathis}\ and\ \citenamefont {Kastner}(1987)}]{Stathis87}%
  \BibitemOpen
  \bibfield  {author} {\bibinfo {author} {\bibfnamefont {J.~H.}\ \bibnamefont {Stathis}}\ and\ \bibinfo {author} {\bibfnamefont {M.~A.}\ \bibnamefont {Kastner}},\ }\href {\doibase 10.1103/physrevb.35.2972} {\bibfield  {journal} {\bibinfo  {journal} {Physical Review B}\ }\textbf {\bibinfo {volume} {35}},\ \bibinfo {pages} {2972} (\bibinfo {year} {1987})}\BibitemShut {NoStop}%
\bibitem [{\citenamefont {Johnson}\ and\ \citenamefont {Samwer}(2005)}]{Johnson05}%
  \BibitemOpen
  \bibfield  {author} {\bibinfo {author} {\bibfnamefont {W.~L.}\ \bibnamefont {Johnson}}\ and\ \bibinfo {author} {\bibfnamefont {K.}~\bibnamefont {Samwer}},\ }\href {\doibase 10.1103/physrevlett.95.195501} {\bibfield  {journal} {\bibinfo  {journal} {Physical Review Letters}\ }\textbf {\bibinfo {volume} {95}},\ \bibinfo {pages} {195501} (\bibinfo {year} {2005})}\BibitemShut {NoStop}%
\bibitem [{\citenamefont {Dieterich}(1979)}]{Dieterich79}%
  \BibitemOpen
  \bibfield  {author} {\bibinfo {author} {\bibfnamefont {J.~H.}\ \bibnamefont {Dieterich}},\ }\href {\doibase 10.1029/jb084ib05p02161} {\bibfield  {journal} {\bibinfo  {journal} {Journal of Geophysical Research: Solid Earth}\ }\textbf {\bibinfo {volume} {84}},\ \bibinfo {pages} {2161} (\bibinfo {year} {1979})}\BibitemShut {NoStop}%
\bibitem [{\citenamefont {Hartley}\ and\ \citenamefont {Behringer}(2003)}]{Hartley03}%
  \BibitemOpen
  \bibfield  {author} {\bibinfo {author} {\bibfnamefont {R.~R.}\ \bibnamefont {Hartley}}\ and\ \bibinfo {author} {\bibfnamefont {R.~P.}\ \bibnamefont {Behringer}},\ }\href {\doibase 10.1038/nature01394} {\bibfield  {journal} {\bibinfo  {journal} {Nature}\ }\textbf {\bibinfo {volume} {421}},\ \bibinfo {pages} {928} (\bibinfo {year} {2003})}\BibitemShut {NoStop}%
\bibitem [{\citenamefont {Wisitsorasak}\ and\ \citenamefont {Wolynes}(2017)}]{Wisitsorasak17}%
  \BibitemOpen
  \bibfield  {author} {\bibinfo {author} {\bibfnamefont {A.}~\bibnamefont {Wisitsorasak}}\ and\ \bibinfo {author} {\bibfnamefont {P.~G.}\ \bibnamefont {Wolynes}},\ }\href {\doibase 10.1073/pnas.1620399114} {\bibfield  {journal} {\bibinfo  {journal} {Proceedings of the National Academy of Sciences}\ }\textbf {\bibinfo {volume} {114}},\ \bibinfo {pages} {1287} (\bibinfo {year} {2017})}\BibitemShut {NoStop}%
\bibitem [{\citenamefont {Cao}\ \emph {et~al.}(2019)\citenamefont {Cao}, \citenamefont {Short},\ and\ \citenamefont {Yip}}]{Cao19}%
  \BibitemOpen
  \bibfield  {author} {\bibinfo {author} {\bibfnamefont {P.}~\bibnamefont {Cao}}, \bibinfo {author} {\bibfnamefont {M.~P.}\ \bibnamefont {Short}}, \ and\ \bibinfo {author} {\bibfnamefont {S.}~\bibnamefont {Yip}},\ }\href {\doibase 10.1073/pnas.1907317116} {\bibfield  {journal} {\bibinfo  {journal} {Proceedings of the National Academy of Sciences}\ }\textbf {\bibinfo {volume} {116}},\ \bibinfo {pages} {18790} (\bibinfo {year} {2019})}\BibitemShut {NoStop}%
\end{thebibliography}%

See Supplemental Material for details on: 1) sample composition, fabrication and basic mechanical propoerties 2) micropillar fabrication by RIE and FIB 3) structural characterisation by micro-Raman spectroscopy 4) relaxation measurements under electron irradiation.

\end{document}